# POST-HOC EXPLAINABILITY OF BI-RADS DESCRIPTORS IN A MULTI-TASK FRAMEWORK FOR BREAST CANCER DETECTION AND SEGMENTATION

*Mohammad Karimzadeh, Aleksandar Vakanski, Min Xian, Boyu Zhang*

Department of Computer Science, University of Idaho, Idaho Falls, ID 83402, USA

## ABSTRACT

Despite recent medical advancements, breast cancer remains one of the most prevalent and deadly diseases among women. Although machine learning-based Computer-Aided Diagnosis (CAD) systems have shown potential to assist radiologists in analyzing medical images, the opaque nature of the best-performing CAD systems has raised concerns about their trustworthiness and interpretability. This paper proposes MT-BI-RADS, a novel explainable deep learning approach for tumor detection in Breast Ultrasound (BUS) images. The approach offers three levels of explanations to enable radiologists to comprehend the decision-making process in predicting tumor malignancy. Firstly, the proposed model outputs the BI-RADS categories used for BUS image analysis by radiologists. Secondly, the model employs multi-task learning to concurrently segment regions in images that correspond to tumors. Thirdly, the proposed approach outputs quantified contributions of each BI-RADS descriptor toward predicting the benign or malignant class using post-hoc explanations with Shapley Values.

## 1. INTRODUCTION

Breast cancer is a devastating disease that affects millions of women worldwide [1], and therefore, development of accurate diagnosis systems is crucial for improving patient outcomes. Medical imaging, including mammography, MRI, and ultrasound imaging, plays a critical role in the early detection and diagnosis of breast cancer [2]. Breast Ultrasound (BUS) has several important advantages in comparison to the other imaging modalities, since it is a non-invasive imaging technique that uses sound waves to create images of breast tissue [3].

The use of machine learning (ML) algorithms for analysis of medical images has shown potential for improved accuracy of the CAD systems for breast cancer detection [4, 5]. However, healthcare is a high-risk domain, and healthcareprofessionals have yet to widely embrace ML-based models in their daily work. The lack of transparency and interpretability remains one of the primary obstacles to the adoption of ML approaches in clinical practice.

To overcome such challenges, explainable AI (XAI) techniques were developed that enable humans to understand and interpret the predictions by AI methods. An important category of XAI techniques involves post-hoc visualizations (e.g., based on CAM [6], Grad-CAM [7]), that use saliency maps, heatmaps, and other forms to provide visual explanations of a previously trained model. The objective is to highlight the regions of the image that contributed the most to the classification decision. However, because a specific region of an image can contain multiple patterns, post-hoc visualizations cannot confirm whether all relevant patterns for explainability have been captured. Conversely, ad-hoc explainable methods provide direct explanations, either by virtue of the structure or the decision-making process of the model. Still, such inherently explainable models can compromise the accuracy of the predictions, which introduces a trade-off between explainability and accuracy.

In this paper, we introduce a novel explainable deep learning (DL) model for breast cancer detection in BUS images, MT-BI-RADS, which uses a hybrid ad-hoc and post-hoc explainability approach. The proposed method provides both visual and quantitative explanations. Specifically, the model outputs the predicted category of each BI-RADS descriptor, as well as, provides visual explanations via highlighting tumor regions in the images by employing image segmentation. In addition, SHAP technique [8] is applied to provide post-hoc quantitative explanations of the significance of each BI-RADS descriptor toward the model predictions. Experimental validation using a dataset of BUS images demonstrated comparatively high accuracy, sensitivity, and specificity of the proposed model. Our approach aims to enhance human-aided diagnosis by providing information to assist clinicians in understanding the predicted tumor class and verifying the accuracy of the CAD system's predictions.

Although previous works in the literature have introduced DL architectures for predicting the BI-RADS descriptors in BUS images [9, 10], the majority of the works did not employ XAI techniques, making it challenging to understand how the models arrived at the conclusions. Several works made efforts to bridge this gap by introducing explainability techniques. The closest to ours is the approach by Zhang et al. [11] referred to as BI-RADS-Net, which introduced an ad-hoc explainability approach to output the category of each BI-RADS descriptor and the probability of malignancy of masses in BUS images. Our proposed approach MT-BIRADS builds upon similar concepts introduced in BI-RADS-Net, and extends it by employing image segmentation techniques and post-hoc explanations to provide important additional insights regarding the decision-making process of the model.

The main contributions of the proposed work include:
- Development of an ad-hoc explainable multitask learning approach for breast cancer detection that concurrently outputs BI-RADS descriptors, tumor segmentation masks, and the tumor class.
- Improvement in the predictive abilities of the model by incorporating branches for BI-RADS descriptors classification and tumor segmentation into a multi-task learning framework.
- Providing post-hoc explanations using Shapley Values to quantify the contributions of the BI-RADS descriptors to the tumor classification into benign or malignant categories.

## 2. RELATED WORKS

Breast Ultrasound is a safe and non-invasive imaging technique that uses high-frequency sound waves to produce detailed images of the internal structures of the breast. Recent progress in ML-based image processing resulted in a large body of work for analysis of BUS images. Prevalent has been the DL methods for image classification based on Convolutional Neural Networks, such as AlexNet, DenseNet, VGG, ResNets, etc. [4, 5]. Recent works also introduced various attention mechanisms [12] and Vision Transformers [13] for this task. Furthermore, numerous works proposed ML models for breast tumor segmentation, among which the U-Net architecture [14] has been the most commonly used.

Several related studies introduced XAI methods for breast cancer detection based on the BI-RADS lexicon. For example, Shen et al. [15] introduced an explainable ML classifier for locating suspected lesions in mammograms. Wu et al. [16] proposed DeepMiner, a DL architecture for tumor classification that uses BI-RADS descriptors for generating text explanations in mammography. Also, Kim et al. [17] developed DL models that utilized the tumor shape and margin in mammograms to predict the class label and BI-RADS category. A major drawback of these approaches in mammography is that they rely on two or three BI-RADS descriptors, which may not provide enough information to fully elucidate the complex process of tumor classification.

Among the XAI methods for BUS images, Zhang et al. [18] employed only the shape and margin BI-RADS descriptors to predict the tumor class. In [19], an ensemble model was proposed with explanations based on statistical texture features of BUS images, which are less useful for radiologists. Additionally, approaches that concentrated on generating textual reports for explaining NN models for BUS [20] were proposed in the literature. BI-RADS-Net [11] proposed an ad-hoc explainable DL model for tumor classification. Despite the advancements from the efforts in related works, explainability of CAD systems for breast cancer diagnosis is still an open research problem that requires further attention.

## 3. PROPOSED METHOD

### 3.1. BI-RADS Lexicon

The BI-RADS system developed by the American College of Radiology has played a pivotal role in standardizing breast imaging assessment and reporting for mammography, ultrasound, and MRI. The system provides a comprehensive lexicon for mass findings categorized into the seven assessment categories presented in Table I. The assessment categories range from 0 to 6, indicating the likelihood of malignancy on a scale of 0% to 100%. These categories are crucial for cancer risk management, where patients categorized as 0 to 3 category undergo short-term follow-up imaging, while those categorized as 4 and 5 undergo diagnostic biopsy.

In addition to the assessment categories, the BI-RADS lexicon also provides standardized terminology for describing various features of mass findings in BUS, including shape, orientation, margin, echo pattern, and posterior features. The standardized terms of the descriptors and their respective classes can be found in Table II. The BI-RADS system serves as a valuable tool in ensuring consistent and accurate breast cancer screening and diagnosis.

### 3.2. Data

The proposed model was trained using 2,186 BUS images obtained by combining three datasets: BUSI [21], BUSIS [22], and HMSS [23]. BUSIS dataset consists of 562 images, of which 306 images contain benign and 256 contain malignant tumors. BUSI dataset includes 630 images that contain mass findings, of which 421 have benign and 209 have malignant tumors. Also, HMSS contains 1,725 images of which 731 have benign and 994 have malignant tumors.

TABLE I. BI-RADS ASSESSMENT CATEGORIES

| Category | Assessment | Likelihood of Malignancy | Management |
|---|---|---|---|
| 0 | Incomplete | N/A | Additional imaging required |
| 1 | Negative | No cancer detected | Annual screening |
| 2 | Benign | 0% | Annual screening |
| 3 | Probably benign | 0-2% | Follow-up in 6 months |
| 4A | Suspicious | 2-10% | Tissue diagnosis |
| 4B | Suspicious | 10-50% | Tissue diagnosis |
| 4C | Suspicious | 50-95% | Tissue diagnosis |
| 5 | Malignant | >95% | Tissue diagnosis |
| 6 | Biopsy-proven malignancy | Cancer present | Surgical excision |

TABLE II. BI-RADS DESCRIPTORS FOR BUS IMAGES

| BI-RADS Descriptors | Descriptors Class |
|---|---|
| Shape | Oval, Round, Irregular |
| Orientation | Parallel, Not parallel |
| Margin | Circumscribed, Not circumscribed (Indistinct, Angular, Microlobulated, Spiculated) |
| Echo Pattern | Anechoic, Hypoechoic, Isoechoic, Hyperechoic, Complex cystic and solid, Heterogeneous |
| Posterior Features | No posterior features, Enhancement, Shadowing, Combined pattern |

The patterns of benign tumors in BUS images are typically characterized by being parallel to the skin surface, oval-shaped, and having a circumscribed margin. In contrast, malignant samples exhibit a more diverse range of appearances and can vary significantly from image to image. Accordingly, learning feature representations of malignant tumors in BUS is more challenging in comparison to benign tumors. Additionally, missing a malignant tumor has more detrimental outcomes in comparison to incorrectly classifying a benign tumor as malignant. Accordingly, for experimental validation, we used solely the malignant samples from the HMSS dataset with the other two datasets.

For more information on BUSI, BUSIS, and HMSS, please refer to their respective sources [21, 22, 23].

### 3.3. Network Architecture

The architecture of MT-BI-RADS is depicted in Fig 1. It consists of a shared backbone network (encoder), a segmentation decoder for reconstructing the tumor masks, and a second branch for predicting the BI-RADS descriptors and tumor class. The encoder employs convolutional and max-polling layers for extracting relevant features in BUS images. The feature maps are utilized by the BI-RADS descriptors branch to predict the descriptors presented in Table II. The resulting predictions are merged and used as an intermediate input to the tumor classification branch, which predicts whether the tumor is benign or malignant. Concurrently, the output of the encoder is fed into the segmentation branch which uses a U-Net-like decoder sub-network, which upsamples the representations of the backbone network and constructs a segmentation mask that identifies the location of the tumor.

Hence, the multitask model includes the following tasks: BI-RADS descriptors (Task 1-5), sub-classes for the margin BI-RADS descriptor (Task 6-9), tumor classification (Task 10), and image segmentation (Task 11). For the classification tasks cross-entropy loss $\mathcal{L}_{CE}$ is utilized, whereas for the segmentation task Dice loss $\mathcal{L}_{Dice}$ is used. The loss function is $\mathcal{L} = \sum_{i=1}^{K} \lambda_i \mathcal{L}_{CE}(X_i, Y_i) + \lambda_{11} \mathcal{L}_{Dice}$, and it is a weighted sum of the losses for the individual tasks where $X_i$ and $Y_i$ are the predicted and ground-truth values, and $\lambda_i$ denote the weight coefficients, respectively.

MT-BI-RADS is an extended version of BI-RADS-Net [11], which has similarity to the upper part in Fig. 1, and contains branches that output the BI-RADS descriptors and the tumor class. However, BI-RADS-Net does not offer information about the tumor locations in BUS images or quantifiable contributions of the BI-RADS descriptors to the tumor class. To address these limitations, MT-BI-RADS introduces a segmentation branch and leverages SHAP values to quantify the contributions of the BI-RADS descriptors.

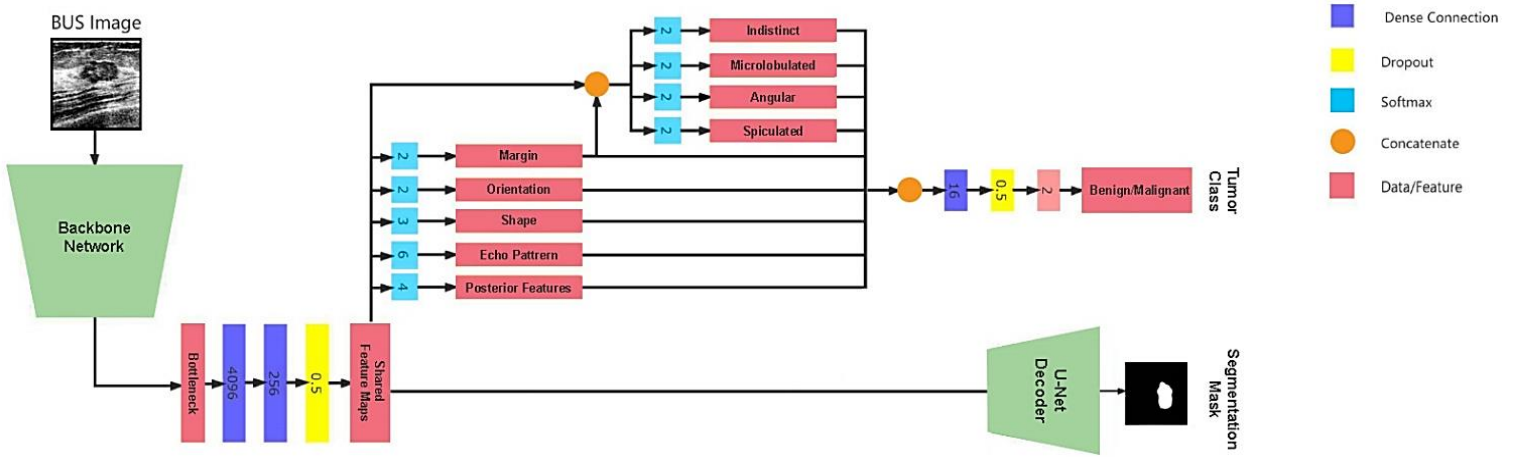

**Fig. 1.** Network architecture of the proposed **MT-BI-RADS** method for BUS analysis. The model consists of a backbone network for feature extraction, shared by a branch for tumor segmentation and a branch for classification of the BI-RADS descriptors and the tumor class (benign or malignant). The tumor segmentation branch combines a VGG-16-like encoder with a corresponding U-Net-like decoder by upsampling the representations from the convolutional layer outputs of VGG-16 and concatenating them with skip connections from VGG-16.

## 3.4. SHAP Method

Our work employs SHAP to determine the contributions by the BI-RADS descriptors to the tumor class prediction. It is based on the concept of Shapley values [24] from cooperative game theory that allocates the payoff of a cooperative game among the players based on their contributions. In the context of ML, Shapley values are used to measure the contribution of each input feature to the output of a model. The Shapley value for a feature represents the average change in the model output when that feature is included in all possible subsets of features, weighted by the number of subsets that include that feature. Shapley values are calculated based on:

$$\phi_i(f) = \sum_{S \subseteq F \setminus \{i\}} \frac{|S|!\,(|F| - |S| - 1)!}{|F|!} [f(S \cup \{i\}) - f(S)]$$

where $\phi_i(f)$ is the Shapley value of feature $i$ in instance $f$. The sum is taken over all subsets $S$ of features that do not include $i$, and $f(S \cup \{i\}) - f(S)$ is the marginal contribution of the feature $i$ when it is added to the subset $S$.

SHAP (SHapley Additive exPlanations) [8] extends the concept of Shapley values for decomposing the prediction of a model into the contributions from each input feature. We used SHAP method as a post-hoc explanation technique for a trained model that assigns contribution values to each BI-RADS descriptor regarding the tumor class. The contribution values allow radiologists to understand which BI-RADS descriptors played the most significant role in predicting the class (benign or malignant) for each BUS image.

## 3.5. Implementation Details

To ensure that important features such as tumor shape and orientation are not distorted in resized images, the original BUS images were first cropped to the largest square segment containing the tumor before being resized to 256×256 pixels. Additionally, to create a more consistent set of images, each single-channel gray BUS image was added to two more channels. One channel was created through histogram equalization, and another by smoothing the gray channel.

The set of images was split into 80% training and 20% testing sets using five-fold cross-validation, with 15% of the images in the training set reserved for validation. The backbone network used for the encoder is a VGG-16 pre-trained on the ImageNet dataset. During training, all layers in the encoder were updated. Data augmentation techniques were applied to the images, including zoom (20%), width shift (10%), rotation (5 degrees), shear (20%), and horizontal flip. A batch size of six images was used, and the models were trained using adaptive moment estimator optimization (Adam) with an initial learning rate of $10^{-5}$, which was reduced to $10^{-6}$ if the loss of the validation set did not decrease for 15 epochs. The training was stopped when the loss of the validation set did not reduce for 20 epochs. For the weight coefficients $\lambda_1$ to $\lambda_{11}$ in the loss, the following values were adopted: 0.2, 0.2, 0.2, 0.2, 0.2, 0.1, 0.1, 0.1, 0.1, 0.5, 0.6, assigned to the orientation, shape, margin, echo pattern, posterior features, indistinct margin, angular margin, spiculated margin, microlobulated margin, tumor classification, and image segmentation branches, respectively. The largest weights were assigned to the tumor classification and segmentation branches.

# 4. EXPERIMENTAL RESULTS

## 4.1. Ad-hoc Explanations

The results from the experimental validation of MT-BI-RADS are shown in Table III. The proposed approach furnishes ad-hoc explainability concurrently with the training/testing phases, by offering explanations regarding the tumor class, BI-RADS descriptors, and segmented tumor region. The model achieved tumor class accuracy of 91.3%, and over 80% accuracy for all five BI-RADS descriptors. Importantly, sensitivity reached 94%. Due to space limitation, the results for the margin sub-classes are not presented in the table, however, for all 4 sub-classes the accuracy exceeded 80%. The table also presents the results of an ablation study performed to evaluate the impact of the different components in the design of MT-BI-RADS. The study assesses the contributions by data augmentation, pretrained network parameters, additional image channels with histogram equalization and smoothing, and cropping of the original images to square-size segments. In addition, the performance metrics for models using different backbone networks (DenseNet12, ResNet50, and MobileNet) and BI-RADS-Net are presented in the table.

TABLE III. ABLATION STUDY OF THE IMPACT OF DIFFERENT COMPONENTS IN THE NETWORK DESIGN ON THE PERFORMANCE

| Method | Tumor Class | | | BI-RADS Descriptors | | | | | Segmentation |
|---|---|---|---|---|---|---|---|---|---|
| | Accuracy | Sensitivity | Specificity | Orientation | Shape | Margin | Echo. Pat | Post. Feat | Dice Score |
| **MT-BI-RADS** | **0.913** | **0.940** | 0.858 | 0.845 | 0.884 | 0.886 | 0.806 | 0.839 | 0.827 |
| Without Augmentation* | 0.892 | 0.897 | 0.880 | 0.850 | **0.887** | 0.875 | 0.808 | 0.835 | 0.813 |
| Without Pretraining* | 0.876 | 0.895 | 0.837 | 0.778 | 0.841 | 0.838 | 0.696 | 0.718 | 0.790 |
| Single Channel Images* | 0.828 | 0.917 | 0.650 | 0.765 | 0.823 | 0.810 | 0.664 | 0.689 | 0.781 |
| Without Image Cropping* | 0.790 | 0.931 | 0.508 | 0.734 | 0.788 | 0.777 | 0.646 | 0.654 | 0.742 |
| DenseNet Backbone | 0.908 | 0.912 | **0.898** | 0.850 | 0.874 | 0.880 | 0.815 | 0.805 | 0.837 |
| ResNet Backbone | 0.903 | 0.925 | 0.856 | **0.864** | 0.882 | 0.891 | **0.814** | 0.820 | **0.842** |
| MobileNet Backbone | 0.909 | 0.930 | 0.869 | 0.848 | 0.885 | 0.807 | 0.796 | **0.867** | 0.841 |
| BI-RADS-Net [11] | 0.900 | 0.923 | 0.885 | 0.848 | 0.877 | **0.887** | **0.814** | 0.834 | N/A |

\* The ablation steps are progressively applied, i.e., the model without augmentation is afterward evaluated without pretrained weights, etc.

## 4.2. Post-hoc Explanations

We considered two approaches for post-hoc explanations of the trained model: SHAP and LIME (Local Interpretable Model-Agnostic Explanations) [25]. The results of the two approaches were comparable for most cases, however for more challenging cases SHAP explanations were more consistent, and consequently, we adopted the SHAP approach. Examples of SHAP explanations for four BUS images are presented in Figs. 2-5.

Fig. 2 displays the predicted segmentation mask and the SHAP values for a benign sample. The values indicate that the probability of the sample being parallel to the skin is 99.87%, having a circumscribed margin is 99.99%, having a hypoechoic echo pattern is 100%, as well as there is a fairly low probability of the sample having lobulated and angular margins. These explanations imply that the model correctly interpreted that all these descriptors led to the benignancy of the prediction, as expected.

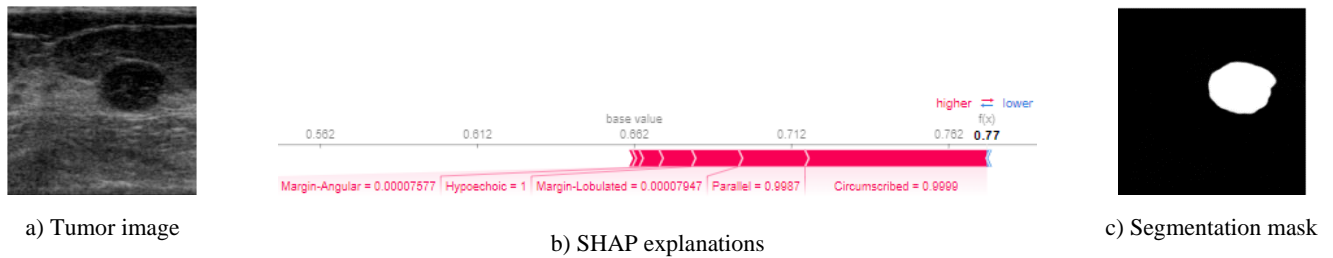

a) Tumor image     b) SHAP explanations     c) Segmentation mask

**Fig. 2.** Image, segmentation mask, and post-hoc explanations of a benign case with parallel orientation, circumscribed margin, and hypoechoic echo pattern.

Fig. 3 displays the output mask of MT-BI-RADS for a benign tumor sample. The SHAP explanations reveal that although the attributes of having a hypoechoic echo pattern and low probability of being oval shaped and low probability of spiculated margin contributed towards the tumor's malignancy (indicated by the blue bar), still the posterior enhancement, being parallel to the skin, and having a circumscribed margin were stronger predictors of benignancy, ultimately leading to being classified as benign.

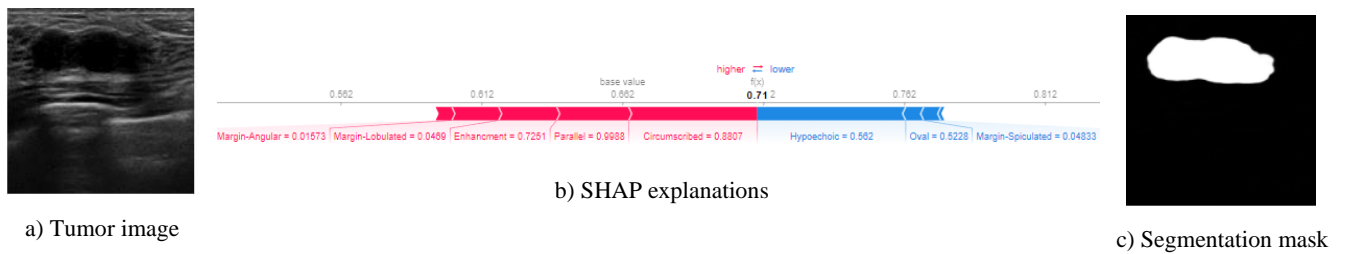

a) Tumor image     b) SHAP explanations     c) Segmentation mask

**Fig. 3.** Image, segmentation mask, and post-hoc explanations of a benign case with a circumscribed margin, parallel orientation, enhancement posterior feature, hypoechoic echo pattern, spiculation, and oval shape, having both malignancy and benignancy signs.

Fig. 4 displays the output mask of the model for a malignant case. According to the SHAP explanations, the spiculated margin, angular margin, not being parallel to the skin, not having a circumscribed margin, and having an irregular shape contributed the most to the malignancy. This is expected, as these are well-known malignancy characteristics of breast tumors. Although the isoechoic echo pattern contributed to the prediction, this is not a typical malignancy sign and requires further analysis.

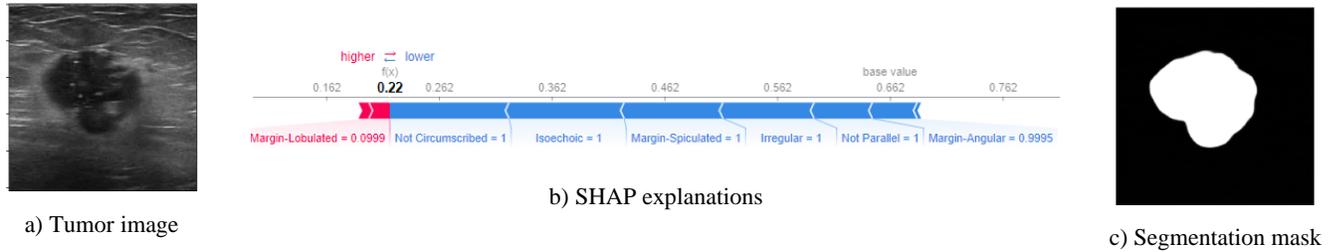

a) Tumor image     b) SHAP explanations     c) Segmentation mask

**Fig. 4.** Image, segmentation mask, and post-hoc explanations of a malignant case with a non-circumscribed margin, isoechoic echo pattern, spiculated and angular margin, irregular shape, and an orientation not parallel to the skin.

Fig. 5 displays the output mask of the model for a malignant case with a non-circumscribed margin, irregular shape, acoustic shadowing, microlobulated margin, angular margin, and hypoechoic echo pattern. SHAP explanations show that these characteristics have significantly contributed to the malignant prediction, whereas being parallel to the skin had small benign contributions. These explanations match well the malignancy characteristics of breast tumors.

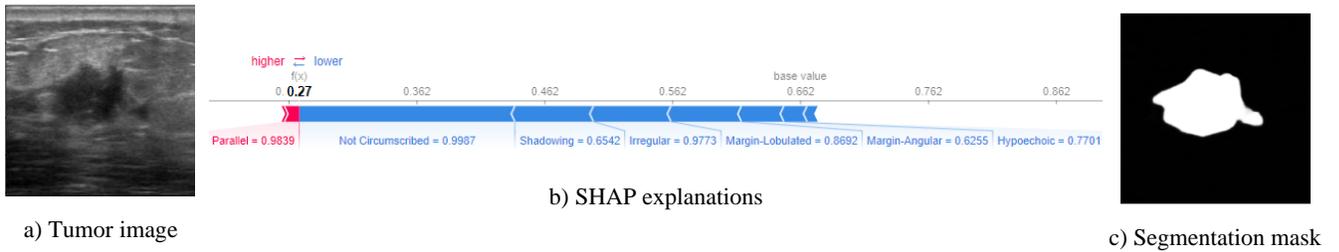

a) Tumor image     b) SHAP explanations     c) Segmentation mask

**Fig. 5.** Image, segmentation mask, and post-hoc explanations of a malignant case with a non-circumscribed margin, shadowing, irregular shape, lobulated and angular margin, hypoechoic echo pattern, an orientation parallel to the skin, and acoustic shadowing.

## 5. CONCLUSION

This paper proposes an explainable DL model for breast cancer detection in BUS images based on BI-RADS descriptors and multitask classification and segmentation learning. The approach uses image segmentation and Shapley Values to provide both ad-hoc and post-hoc explanations for its predictions. MT-BI-RADS achieves high levels of accuracy, sensitivity, and specificity, and provides quantitative and visual explanations of the tumor regions. The proposed approach has the potential to improve the accuracy and interpretability of CAD systems for breast cancer detection, and could serve as a valuable tool for medical professionals in their efforts toward enhanced diagnosis.

## ACKNOWLEDGMENT

Research reported in this publication was supported by the National Institute of General Medical Sciences of the National Institutes of Health under Award Number P20GM104420. The content is solely the responsibility of the authors and does not necessarily represent the official views of the National Institutes of Health.